# Ferroelectric–Superconducting Interaction in Epitaxial $YBa_2Cu_3O_{7-\delta}/BaTiO_3$ films

Louis Oppong-Antwi[1,2*], David. L. Cortie[3*], Qi Zhang[1,4,5], Avi Bendavid[1], Nagarajan Valanoor[5], Wendy Purches[1], Golrokh Akhgar[1*]

*[1] CSIRO Technology, Lindfield, NSW 2070, Australia*
*[2] CSIRO Energy, Newcastle Energy Center, 10 Murray Dwyer Circuit, Mayfield West, NSW 2304, Australia*
*[3] Australian Nuclear Science and Technology Organisation, New Illawarra Road, Lucas Heights, NSW, 2234, Australia*
*[4] School of Engineering, Macquarie University, NSW 2109, Australia*
*[5] School of Materials Science and Engineering, UNSW, Sydney, NSW 2052, Australia*

**Corresponding Authors: louis.oppong-antwi@csiro.au; Golrokh.Akhgar@csiro.au; dcr@ansto.gov.au*

## Abstract

Microscopic theories predict that the critical temperature of a superconducting layer can be strongly modified in proximity to a ferroelectric, however, experimental evidence is lacking. We report on $BaTiO_3$ (BTO)/ $YBa_2Cu_3O_{7-\delta}$ (YBCO) heterostructures grown on $SrTiO_3$ (001) substrates using pulsed laser deposition (PLD) with precise control of growth conditions and oxygen stoichiometry. Piezoresponse force microscopy confirms ferroelectric switching within both single-layer and heterostructured films. Neutron reflectometry measurements show low roughness parameters for the buried interfaces, which is critical for mediating the interfacial coupling between superconductivity and ferroelectric polarization. The superconducting transition temperature ($T_C^R$) measured from the R(T) curve of the single-layer YBCO was $\sim 86$ K and a transition temperature ($T_C^M$) of $\sim 80-85K$ observed from the M(T) curve. For the heterostructure film (YBCO/BTO/YBCO) grown under the same conditions, both resistivity and magnetization measurements indicate a lower $T_c$ ($\sim 75K$ and $\sim 55K$, respectively) and a lower overall susceptibility. These findings highlight the need to combine multiple techniques to characterize BTO/YBCO heterostructures to identify the subtle features in the ferroelectric coupling. This establishes BTO/YBCO heterostructures as promising systems for tunable oxide electronics and reconfigurable superconducting devices.

## Introduction

Complex oxides represent a broad class of materials that display an array of collective electronic phenomena, including superconductivity, ferroelectricity, ferromagnetism, and magnetoresistance, arising from the strong coupling between their charge, spin, orbital, and lattice degrees of freedom [1,2]. When integrated into epitaxial heterostructures, these interactions can be further engineered: strain, interfacial charge transfer, and polar discontinuities at interfaces may induce emergent properties that are not present in bulk phases [3,4].

An insufficiently explored avenue is the coupling of a high-temperature superconductor (SC) with a ferroelectric (FE) oxide, which allows the interplay between a controllable ferroelectric polarization and superconductivity. Such coupling enables dynamically modulated superconducting transport via electric fields, providing a foundation for novel device concepts such as superconducting field-effect transistors, non-volatile memories, and electrostatically gated Josephson junctions.

Theoretically, several coupling mechanisms have been proposed: electron-polarization coupling that modulates interfacial carrier density [5], pairing symmetry changes near ferroelectric quantum critical points [6], and size-induced $T_c$ and magnetic-critical-field ($H_c$) suppression in coupled FE-superconducting domains [7]. According to theory, the critical temperature of a FE/SC heterostructure depends on the interfacial coupling parameter ($\gamma$) in a nontrivial way: for large $\gamma$, superconductivity can be completely suppressed; for intermediate $\gamma$, the transition temperature is reduced ($0 < T_c < T_c^{\mathrm{bulk}}$); and for sufficiently small $\gamma$, an *increase* in $T_c$ is predicted [8]. Experimental evidence, however, remains divided. Previous studies combining $\mathrm{YBa_2Cu_3O_{7-\delta}}$ with ferroelectric materials such as $\mathrm{La_{0.7}Sr_{0.3}MnO_3}(LSMO)$, $\mathrm{Pb(Zr,Ti)O_3}(PZT)$, $\mathrm{BiFeO_3}(BFO)$ have demonstrated that ferroelectric polarization can effectively modulate the superconducting characteristics of YBCO [9–12].

Additionally, some studies have also reported a negligible change in $\mathrm{T}_c$ [13]. To the best of our knowledge, no experiment has yet demonstrated a clear enhancement of $T_c$ This highlights the importance of understanding the microscopic origins of $\gamma$, particularly the buried interface structure, which has rarely been characterized in prior work despite its relevance to the coupling. Additional factors such as strain, band bending, oxygen-vacancy redistribution, and interfacial disorder further complicate the underlying physics, underscoring that ferroelectric–superconducting coupling is inherently multichannel and requires multiprobe characterization. Most previous studies of ferroelectric–superconducting heterostructures have relied primarily on resistivity measurements to infer superconductivity; however, zero resistance mainly signals percolative connectivity rather than true volume-averaged superconductivity and can obscure interfacial coupling effects. In the presence of strain, polarization-induced carrier modulation, or interfacial disorder, transport measurements alone may underestimate local $T_c$ suppression or spatial inhomogeneity of the superconducting condensate. Combined transport and magnetization measurements are therefore essential.

In this Letter, we report on the synthesis of high-quality $\mathrm{BaTiO_3}$/YBCO heterostructures, studied using a range of experimental techniques including resistivity, magnetometry, and neutron reflectometry to determine their superconducting and interfacial properties. YBCO remains the archetypal high-$T_c$ cuprate superconductor, exhibiting a transition temperature near 90 K in optimally oxygenated films, low microwave surface resistance, and excellent structural compatibility with perovskite oxides.

In our work, $\mathrm{BaTiO_3}$(BTO) was selected as the ferroelectric layer owing to its well-established ferroelectric properties, stable tetragonal perovskite structure, high Curie temperature (120°C) and large spontaneous polarization of $\sim 26\ \mu\mathrm{Ccm}^{-2}$ [14,15]. Its lattice parameters (a = 3.99 Å, c = 4.04 Å [16]) yield a small mismatch, $\sim$ 2.5 % with orthorhombic $\mathrm{YBa_2Cu_3O_{7-\delta}}$(YBCO), enabling high-quality epitaxial growth. BTO also possesses a high dielectric constant ($>$ 1500 in bulk at room temperature, $\sim$ 200 at 77 K) and typically forms smooth, well-ordered thin films, both favorable for integration with perovskite superconductors. Compared to other FE systems, BTO offers distinct advantages: a closer lattice match to YBCO (3.82–3.89 Å vs. 3.99 Å)[17], the absence of magnetic ordering that could influence superconducting coherence, and greater thermal and chemical stability during processing. Epitaxial YBCO/BTO bilayers fabricated by pulsed laser deposition (PLD) on $SrTiO_3$ substrates have been shown to exhibit high crystalline quality and sharp superconducting transitions [13]. Past work for YBCO grown on BTO reported a zero-resistance transition temperature around 88–89 K with narrow transition widths ($\Delta T_c \approx 1.8$ K), while BTO layers retain their ferroelectric functionality, showing dielectric constants near 180 and remanent polarizations of $\sim 3.5\ \mu$C / cm$^2$ at 77 K [13]. These findings underscore the structural and functional compatibility of YBCO and BTO, however the lack of a modified $T_c$ indicates weak $\gamma$ coupling, and further work is required to identify methods to enhance this to enable reconfigurable superconducting electronics.

## Experimental Section

Single-layer and heterostructure thin films were grown on $SrTiO_3$ (STO) (001) substrates using a pulsed laser deposition (PLD) system equipped with a KrF excimer laser ($\lambda = 248nm$) operating at a repetition rate of 5 Hz. The deposition parameters for the YBCO were optimized to achieve a balance between the highest $T_c$ and minimal surface roughness. YBCO films, approximately 150 nm thick, exhibited optimal properties when deposited at a substrate temperature of 780 °C under an oxygen pressure of 200 mTorr. Subsequently, the BTO growth conditions were adjusted to remain as consistent as possible with those of YBCO. BTO film with an average thickness of 90 nm was deposited at 750 °C in 120 mTorr of oxygen. YBCO/BTO/YBCO heterostructures (denoted as YBY) were fabricated using the same optimized parameters as for the individual layers, with the top YBCO layer deposited at similar deposition temperature to the underlying BTO.

## Results and Discussion

The X-ray diffraction (XRD) patterns of the YBCO, BTO, and YBCO/BTO/YBCO heterostructures (Fig. 1a-c) confirm that both layers are highly c-axis oriented, indicating excellent crystallinity and epitaxial alignment with the STO substrate. No secondary phases or impurity peaks were detected, suggesting negligible interdiffusion or interfacial reactions. The out-of-plane lattice parameter of YBCO was determined from the (005) peak of its XRD pattern to be 11.691 Å, slightly larger than the bulk value (11.68 Å), implying a minor lattice strain induced by epitaxial growth [18].

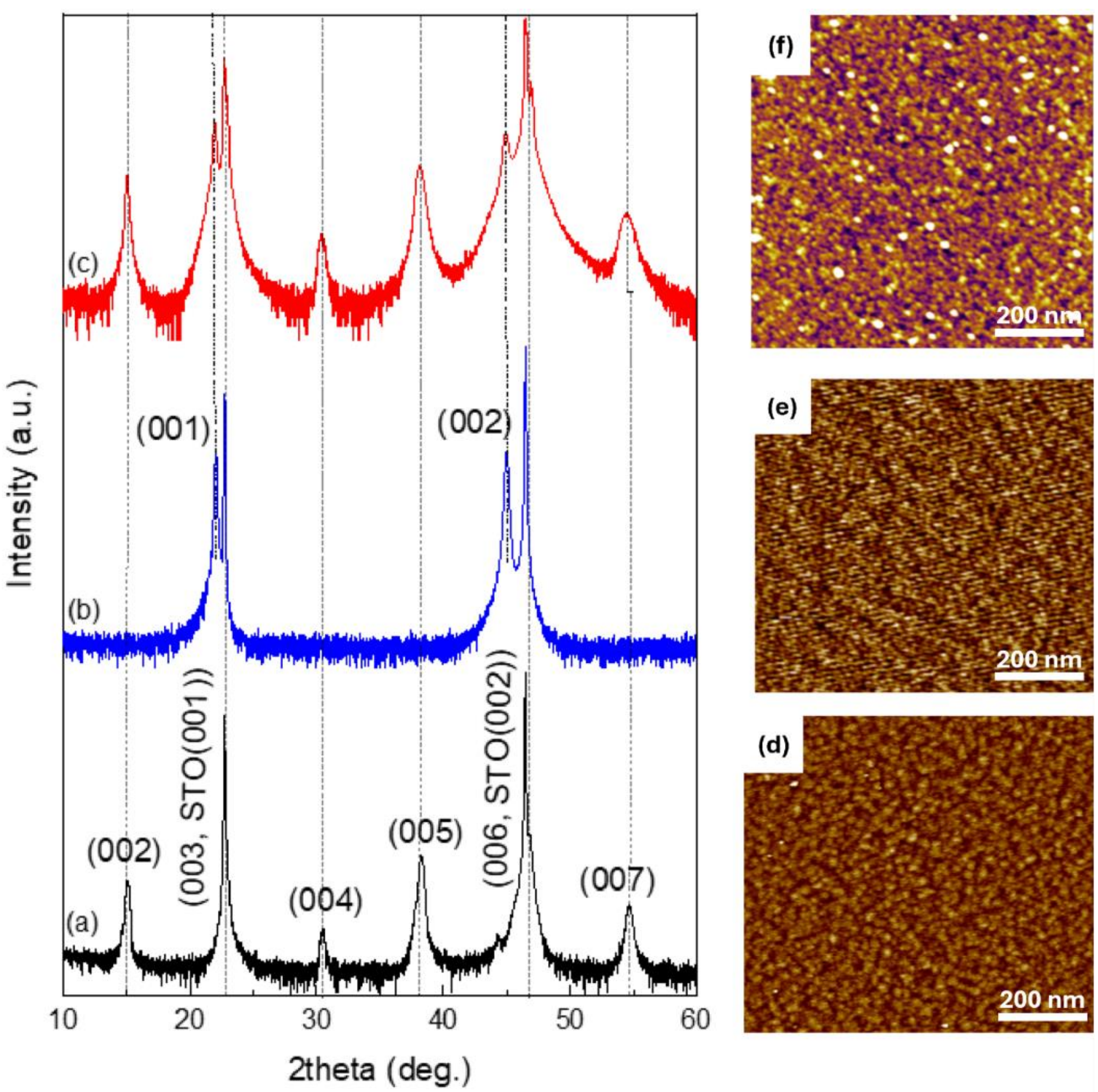

*Fig. 1 The XRD patterns of samples, (a) YBCO, (b) BTO and (c) YBCO/BTO/YBCO films and AFM image of (d) YBCO, (e) BTO and (f) YBCO/BTO/YBCO. Each AFM image of a 2 µm × 2 µm area.*

The c-axis lattice parameter of BTO was 4.03 Å, comparable to bulk BTO [14,19,20], which is normally within the range commonly seen for epitaxial thin films, indicating a biaxial compressive strain. This is in agreement with previous reports of BTO grown on STO substrates[21]. However, for the YBY film, the out-of-plane lattice parameter of YBCO further increases to 11.713 Å, while that of BTO remains similar to that of bulk BTO (4.04 Å).

Atomic force microscopy (AFM) was employed to examine the surface morphology of the YBCO, BTO, and YBCO/BTO/YBCO heterostructure shown in Fig 1 (d-f). The single-layer YBCO films exhibited uniform nanoscale islands, with an average root-mean-square (RMS) roughness of approximately 1.13 nm (Fig 1 (d)) over a 2 × 2 µm² scan area. The observed surface morphology has been reported previously for YBCO thin films due to the strain relaxation during deposition[22]. The BTO film on the other hand showed much smoother surface compared to YBCO, with an average root-mean-square (RMS) of 0.49 nm, seen in Fig 1(e). The YBCO/BTO/YBCO trilayer structure led to a rougher film compared to the individual films. The top surface of the heterostructure becomes rougher with an increased rms of 2.31 nm. The corresponding topographic image in Fig. 1(f) shows grains larger than those on the single YBCO layer. These AFM observations are consistent with the XRD results, further supporting the high crystalline quality and epitaxial integrity of the heterostructures.

To evaluate the intrinsic ferroelectric properties of the BTO thin film, a reference sample was grown under identical deposition conditions on conductive Nb-doped $SrTiO_3$ (Nb-STO, 0.5 wt% Nb), which served as the bottom electrode. Ferroelectric domain behavior was characterized at room temperature using piezo response force microscopy (PFM), enabling simultaneous imaging and local polarization switching. By applying a DC bias between the conductive PFM tip (top electrode) and the Nb-STO substrate, out-of-plane polarization was reversibly switched. An initial downward polarization state could be inverted by applying an electric field exceeding the coercive threshold, with local switching typically achieved at biases of approximately ±2 V.

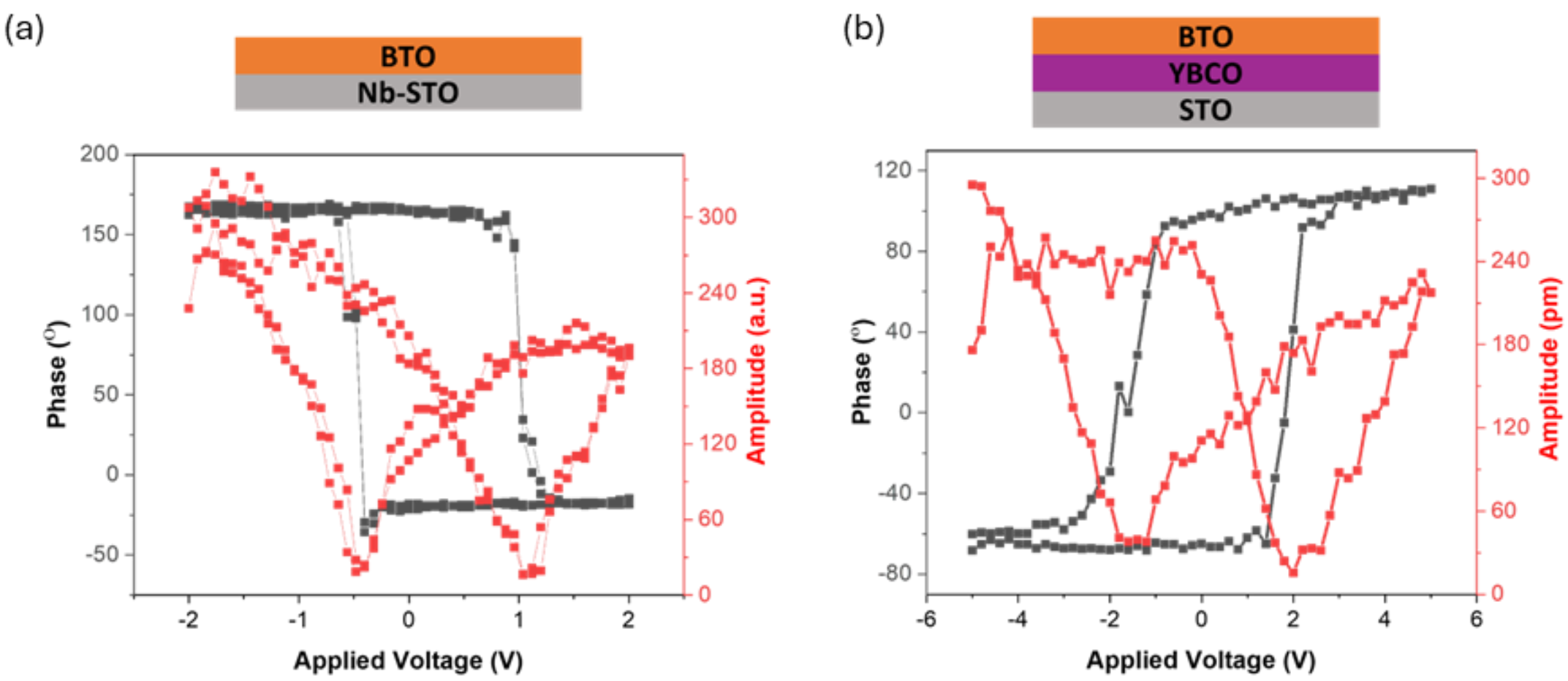

*Fig. 2 Local PFM amplitude (red) and phase (black) hysteresis curve of (a) BTO/nb-STO, (b) YBCO/BTO film*

To probe ferroelectricity within the heterostructure, a BTO/YBCO bilayer was fabricated on STO under the same growth conditions, omitting the top YBCO layer so that the underlying YBCO served as the bottom electrode. PFM measurements confirm robust nanoscale ferroelectricity in both the BTO/Nb-STO reference and the YBCO/BTO heterostructure, as evidenced by clear 180° phase contrast reversal and well-defined butterfly-shaped amplitude loops. However, the YBCO/BTO sample exhibits a higher coercive voltage and reduced piezo response amplitude. This modified switching behavior is attributed to partial oxygen deficiency in the YBCO bottom electrode, as indicated by XRD analysis, which reduces carrier density and weakens interfacial charge screening [23] . The resulting enhanced depolarization fields increase the energy barrier for domain wall motion, broadening the switching loops. Nevertheless, the BTO layer retains clear and reversible ferroelectric functionality within the heterostructure, highlighting the critical role of electrode properties and interfacial electrostatics in governing polarization dynamics.

Neutron reflectometry (NR) measurements were performed on the multilayered YBCO/BTO/YBCO (YBY) films at 120 K using the Platypus reflectometer[24]. As shown in Fig. 3a, the NR data confirm high-quality epitaxial growth with sharp interfaces with the structure shown by the scattering length density profiles in Fig. 3a. The thicknesses and SLD values are as expected YBCO (~ 140 nm, SLD = $4.6 \times 10^{-4}$ Å$^{-2}$), BTO ~80 nm, SLD = $3.2 \times 10^{-4}$ Å$^{-2}$ and top YBCO () ~150 nm

.

*Fig. 3 (a) Neutron reflectometry curve of YBCO and YBCO/BTO/YBCO, (b) corresponding nuclear scattering length density (SLD) depth profiles of the two films*

The reflectivity curves exhibit clear Kiessig fringes to $Q \approx 0.045$ Å$^{-1}$, indicating low surface roughness (1.2 nm). The fitted SLD values for the bottom YBCO layer is $4.63 \pm 0.02$ Å$^{-2}$, close to the values of optimally doped YBCO ($\delta \approx 0.1$). For the top layer grown on BTO the SLD is marginally reduced to $4.52 \pm 0.03$ Å$^{-2}$, implying a larger ($\delta \approx 0.45$), in agreement with the XRD. In the heterostructures, the buried FE/SC interface roughnesses of $1.9 \pm 2.8$ nm and $3.7 \pm 0.9$ nm were determined for the bottom and top interfaces respectively. This may reflect moderate diffusion, and strain relaxation arising from thermal expansion mismatch during cooldown. The neutron reflectometry shows that the film quality is close to ideal, although the moderate roughness and stoichiometry variation are expected to influence the resulting superconducting properties.

The superconducting properties of the single-layer YBCO film were evaluated by measuring the temperature dependence of its electrical resistance under zero magnetic field. As shown in Fig. 4 (black line), the film exhibits a resistive superconducting transition temperature ($T_c^R$) of ~ 86 K, with an onset $T_c^R \approx 88 - 90$ K and a sharp transition width ($\Delta T_c^R \approx 2$ K). Above 90 K, the film displays the metallic behavior characteristic of optimally oxygenated YBCO in the normal state. In comparison, the superconducting transition of the YBCO/BTO/YBCO heterostructure occurs at a lower temperature of ~ 75 K with an onset $T_c^R \approx 84 - 86$ K (Fig. 4 (red line)).

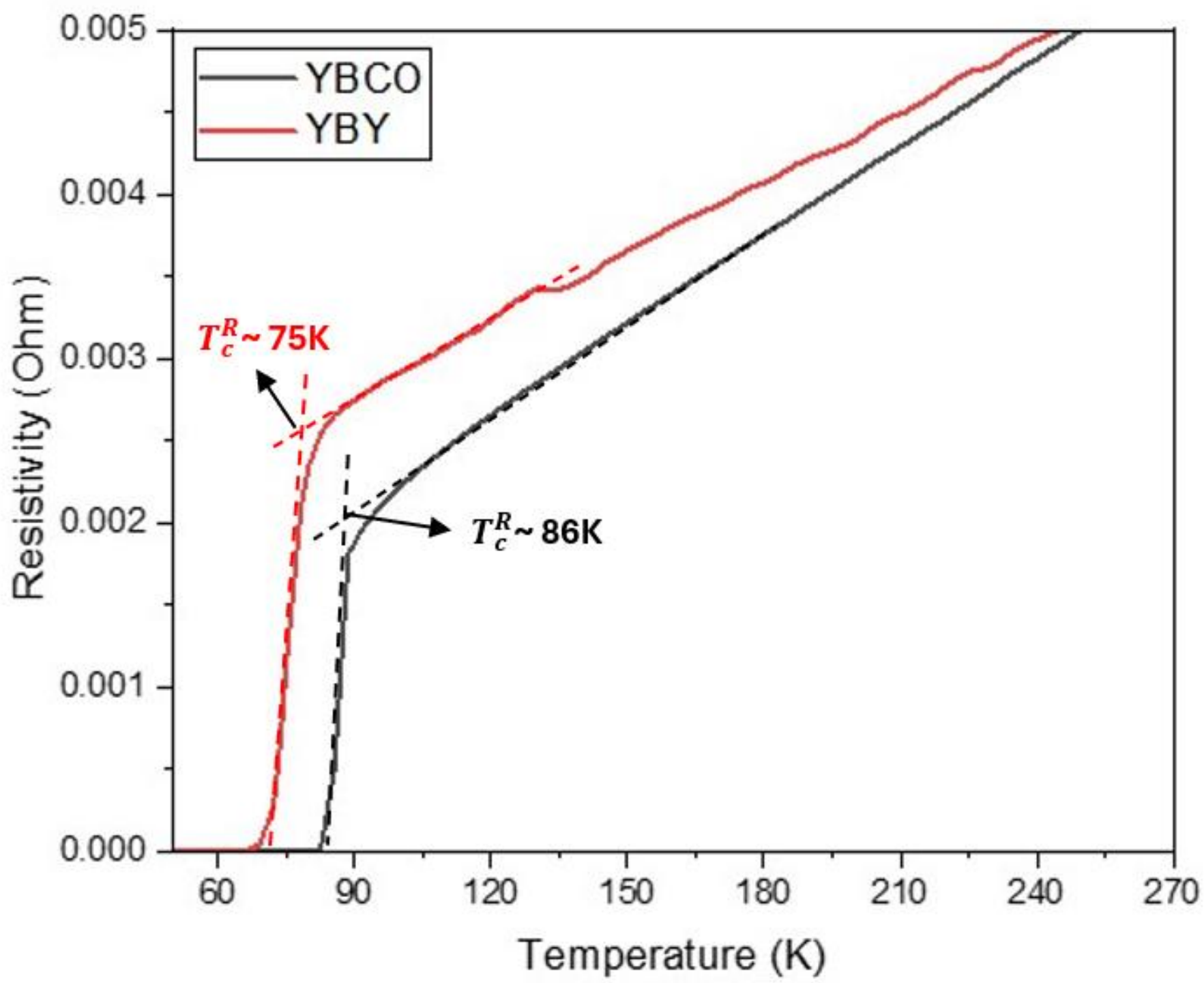

*Fig. 4 Resistance versus temperature of YBCO (black line) and YBCO/BTO/YBCO (YBY)(red line) films*

It is well established that the superconducting transition temperature ($T_c$) of YBCO is closely correlated with the c-axis lattice parameter, which reflects the oxygen content. In oxygen-deficient YBCO films, $T_c$ decreases as oxygen stoichiometry is reduced, accompanied by an expansion of the c-axis [3,25–27]. The reduction in $T_c$ in the heterostructure can be correlated with the observed increase in the out-of-plane lattice parameter from the XRD data, suggesting an increase in the oxygen deficiency parameter $\delta$ in $\mathrm{YBa_2Cu_3O_{7-\delta}}$ from 0 to approximately 0.5 [2,28,29]. This slight elongation of the c-axis is attributed to strain relaxation effects at the interface, because of oxygen depletion and consequently a decrease in $T_c^R$ onset, consistent with the trends reported in previous studies [30,31]. The YBCO films in the heterostructure remained superconducting, but their properties were partially suppressed by interfacial, stoichiometric, and electrostatic effects arising from BTO integration.

In Fig. 5(a) and (b), we also show the M(T) curves under zero field cooled (ZFC) process of YBCO and YBY films, respectively. The onset $T_c$ observed from the M(T) curve for the YBCO was like that of the R(T) curve for the single layer YBCO film, with a superconducting transition temperature $T_c^M \approx 80 - 85$ K.

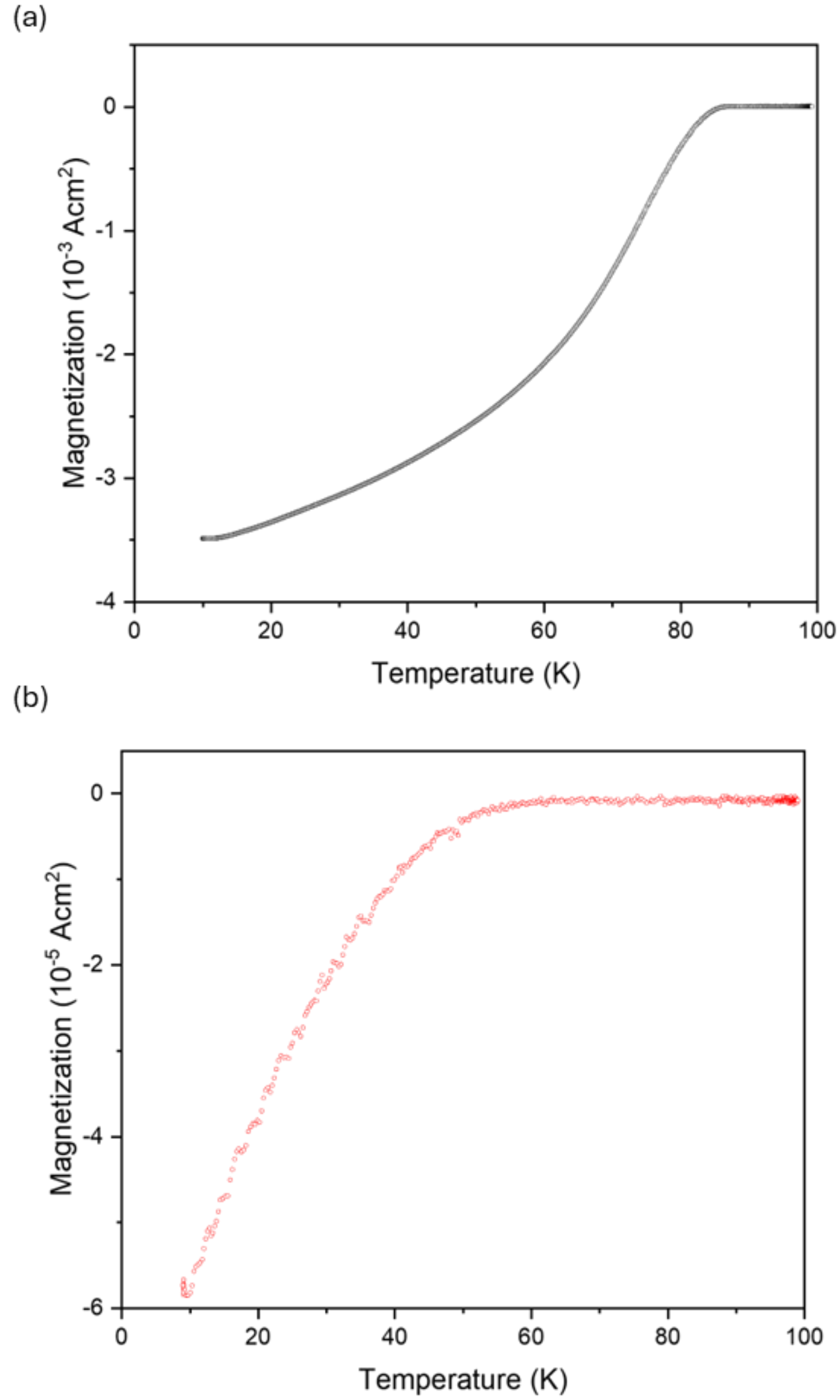

*Fig. 5 ZFC M(T) curves (a)YBCO), and (b) YBCO/BTO/YBCO films measured under an applied field of 20 Oe.*

In the YBY heterostructure, the onset $T_c^M$ extracted from ZFC magnetization is substantially shifted to lower values with a lower $T_c^M$ compared to the R(T) curve of the same film. This significant difference between $T_c$ from the R(T) curve and that of the ZFC magnetization of the YBY film, reflects BTO-induced progressive decoupling of superconducting YBCO grains and inhomogeneous superconducting state arising from interface effects and grain boundary modifications[32].

In such systems, the resistive transition probes the formation of a continuous superconducting percolation path, leading to zero resistivity once the superconducting volume fraction $f_s(T)$ exceeds the critical percolation threshold ($f_c \approx 0.5$–$0.6$) in two- and three-dimensional networks, $\rho(T) = 0$ once $f_s(T) > f_c$. Consequently, zero resistance can be achieved at higher temperatures when a minority fraction of higher-$T_c$ regions forms a connected current-carrying network[33].

In contrast, the ZFC magnetization reflects a volume-averaged diamagnetic response,

$$M(T) \propto \int \chi\text{loc}(\text{T})\, \text{dV} \quad (1)$$

M(T) is the total measured magnetization (magnetic moment), $\chi\text{loc}(\text{T})$ is the local magnetic susceptibility at each point/volume element in the inhomogeneous sample and dV is the differential volume element, which is dominated by the larger fraction of YBCO regions exhibiting reduced local $T_c$. These lower-$T_c$ regions contribute weakly to magnetic shielding near the resistive transition, resulting in a substantially lower apparent transition temperature in magnetization measurements and a broadened superconducting transition.

Interface effects at the BTO/YBCO boundaries, including epitaxial strain, oxygen redistribution, and structural defects, are known to suppress superconductivity in cuprates [34,35] and thus may contribute to the reduced local $T_c$ in the YBY heterostructure. Additionally, the spontaneous ferroelectric polarization of BTO can electrostatically modulate the hole carrier density in adjacent YBCO layers, further altering the superconducting order parameter [5,36]. Within a carrier-density-controlled framework, the superconducting gap amplitude follows;

$$\Delta \propto \sqrt{n_h\left(1 - \frac{n_h}{n_{QCP}}\right)} \quad (2)$$

where spatial variations in the hole concentration $n_h$ induced by interfacial polarization and strain lead to local suppression of superconductivity[37]. Given the intrinsic sensitivity of the $d$-wave order parameter in YBCO to disorder and carrier inhomogeneity, the BTO/YBCO interfaces effectively form Josephson weak links between superconducting regions, broadening the transition and decoupling grains. As a result, superconductivity in the YBY film emerges via percolation of higher-$T_c^R$ regions, while the bulk diamagnetic response measured by ZFC magnetization is governed by the majority fraction of YBCO with reduced $T_c^M$, yielding a large apparent transition width $\Delta$Tc ≈ 20 K[38].

While early work by Hontsu *et al.*[13] demonstrated that epitaxial YBCO/BTO bilayers can retain a bulk-like superconducting transition in transport measurements, the interplay between ferroelectric polarization and superconductivity at the nanoscale has remained insufficiently understood, with conflicting reports regarding the extent of $T_c$ suppression [39–42]. Through a combined transport and magnetization study of epitaxial YBCO/BTO/YBCO trilayers, supported by piezoresponse force microscopy and neutron reflectometry, we observe a pronounced discrepancy between the resistive superconducting transition and the volume-averaged magnetic transition, whereas reference YBCO films show nearly identical transition temperatures in both probes. This divergence indicates the emergence of percolative superconductivity arising from carrier inhomogeneity and interfacial effects associated with the ferroelectric BTO layer. Structural and magnetic analysis indicates that the suppression and broadening of the $T_c$, observed in both transport and ZFC magnetization, arises from: (1) Minimal strain effect ($c_{\text{YBCO}} = 11.71$ Å, $\epsilon_c \approx +0.25\%$, $\Delta T_c < 2$ K); (2) underdoping dominates $\delta \approx 0.45$ ($YBa_2Cu_3O_{6.55}$) from BTO growth oxygen loss; (3) roughness gradient (0.4→1.2 nm) reduces effective top YBCO thickness below coherence length $\xi_N$. Together, these results reconcile longstanding inconsistencies in prior BTO/YBCO studies and establish ferroelectric–superconducting oxide heterostructures as a platform where structurally sharp interfaces can nonetheless host strongly inhomogeneous and percolative superconducting states.

## Conclusion

We have demonstrated the successful epitaxial integration of superconducting YBCO and ferroelectric BTO thin films, achieving high structural quality and preserved ferroelectric functionality. $T_c$ suppression in YBCO/BTO/YBCO reveals ferroelectric-superconducting coupling; combined transport and magnetization distinguish intrinsic effects from percolative transport artifacts. PFM analyses confirm switchable polarization states within the BTO layer, while neutron reflectometry provides further evidence to confirm the interfacial properties observed when these heterostructures are formed. These results collectively reveal that YBCO–BTO heterostructures maintain both superconducting and ferroelectric characteristics, with measurable coupling effects that can be exploited for tunable superconducting devices. Future efforts could focus on in situ electric field gating to directly quantify the polarization-coupling strength and explore the thickness dependence of the ferroelectric layer to decouple strain from electrostatic effects. Additionally, utilizing high-resolution local probes like scanning tunneling spectroscopy could visualize the nanoscale percolative landscapes identified here, providing a pathway toward the practical realization of reconfigurable oxide electronics.

## Author Contribution:

L.O-A, W.P and G.A conceived the project. L.O-A, D.C and G.A carried out the experiment, characterization and wrote the manuscript. L.O-A, D.L.C, G.A and Q.P analyzed data. L.O-A, G.A, W.P, D.L.C, N.V and A.B reviewed the manuscript.

## Acknowledgments:

L.O.A. acknowledges an early career discovery grant from the Australian Institute for Nuclear Science and Engineering (AINSE). The authors acknowledge the Australian Nuclear Science and Technology Organisation (ANSTO) for the facilities. The authors also acknowledge the Solid State and Elemental Analysis Unit (SSEAU) in the Mark Wainwright Analytical Center (MWAC) at the University of New South Wales, Sydney for their support in acquiring the X-ray diffraction data for this study.